\documentclass[conference]{IEEEtran}
\IEEEoverridecommandlockouts
\usepackage[left=1.6cm,right=1.6cm,top=1.6cm, bottom=1.6cm]{geometry}
\usepackage{amsthm}
\usepackage{amssymb}
\usepackage{algorithmic}
\usepackage{graphicx}
\usepackage{hyperref}%
\usepackage{tikz}
\usepackage{textcomp}
\usetikzlibrary{arrows, matrix}
\usepackage{dirtytalk}
\usepackage{verbatim}
\usepackage{amsmath}
\usepackage[ruled,vlined,linesnumbered]{algorithm2e}
\usepackage{array}
\usepackage[T1]{fontenc}
\usepackage{float}
\usepackage{graphicx}
\usepackage{url}
\usepackage{multirow}   
\usepackage{soul}
\usepackage{dblfloatfix}
\usepackage{pgfplots}
\usepackage{multicol}
\usepackage{tcolorbox}
\usepackage[inline]{enumitem}
\usepackage{enumitem}
\usepackage[export]{adjustbox}
\usepackage{float}
\usetikzlibrary{positioning, fit, calc, shapes, arrows}
\usepackage[underline=false]{pgf-umlsd}

\usepackage{graphicx}
\usepackage{caption, balance}
\usepackage{array,amsfonts}
\usepackage{subcaption}
\usepackage{cite}
\usepackage{wrapfig}
\usepackage{tabularx}
\usepackage[english]{babel}	
\usepackage{soul}
\usepackage{float}
\usepackage{tikz}

\usetikzlibrary{fit,positioning,calc}
\tikzset{decision/.style = {diamond, draw, fill=blue!20, text width=4.5em, text badly centered, node
                            distance=3cm, inner sep=0pt},
         block/.style    = {rectangle, draw, fill=black!25, text width=5em, text centered, rounded
                            corners, minimum height=4em},
         line/.style     = {draw, -latex'},
         cloud/.style    = {draw, ellipse,fill=red!20, node distance=3cm, minimum height=2em}
}

\usepackage{epstopdf}
\theoremstyle{remark}

\if CLASSINFOpdf

\else

\fi

\usetikzlibrary{calc}
\usetikzlibrary{shapes.arrows}
\usetikzlibrary{positioning,fit,backgrounds}

\begin{document}

\title{Kisê-Manitow’s Hand in Space: \\ Securing Communication and Connections in Space}
\author{\IEEEauthorblockN{Nesrine Benchoubane\IEEEauthorrefmark{1}\IEEEauthorrefmark{2},
Gunes Karabulut Kurt\IEEEauthorrefmark{1}\IEEEauthorrefmark{2}}
\IEEEauthorblockA{\IEEEauthorrefmark{1,2}Poly-Grames Research Center, Department of Electrical Engineering, Polytechnique Montréal, QC, Canada}
\IEEEauthorblockA{\IEEEauthorrefmark{2}Astrolith, Transdisciplinary Research Unit of Space Resource and Infrastructure Engineering, Polytechnique Montréal, QC, Canada}
\textit{\{nesrine.benchoubane, gunes.kurt\}@polymtl.ca}
}

\maketitle
\begin{abstract}
The lunar Gateway is a complex \textit{system-of-systems} (SoS) requiring secure, resilient integration of all its systems. This paper develops controls for its early systems to secure command and data handling (C\&DH) and evaluates their effectiveness through a case study on collision impact propagation originating from Canadarm3. We demonstrate that a baseline security strategy with stringent, standardized controls reduces failure impact by up to 70\%, underscoring the importance of consistent security integration across all systems. 
\end{abstract}


\IEEEpeerreviewmaketitle


\section{Introduction}

\IEEEPARstart{A}{Nehiyawak} elder once described the land as abundant with life, stressing that new civilizations, despite their technological advancements, must remain in harmony with nature\cite{berkes1999sacred}. He recalled how industrial progress nearly led to the extinction of many species. From this wisdom, two key lessons emerge: \begin{enumerate*}[label=(\roman*)] \item technological advancements must be (and remain) ethical and non-damaging, and \item life, whether on Earth or beyond, will thrive if we listen and adapt responsibly. \end{enumerate*} Rooted in the concept of \textit{Kisê-Manitow}, the Algonquian term for the Great Creator, this understanding acknowledges the importance of ethical and sustainable engineering. Honoring the wisdom of First Nations, the title of this work recognizes Canadarm3 as the \say{Great Creator’s Hand} in space, symbolizing Canada’s latest contributions to space exploration.

\begin{figure}[ht!]
    \centering
    \includegraphics[width=0.8\linewidth]{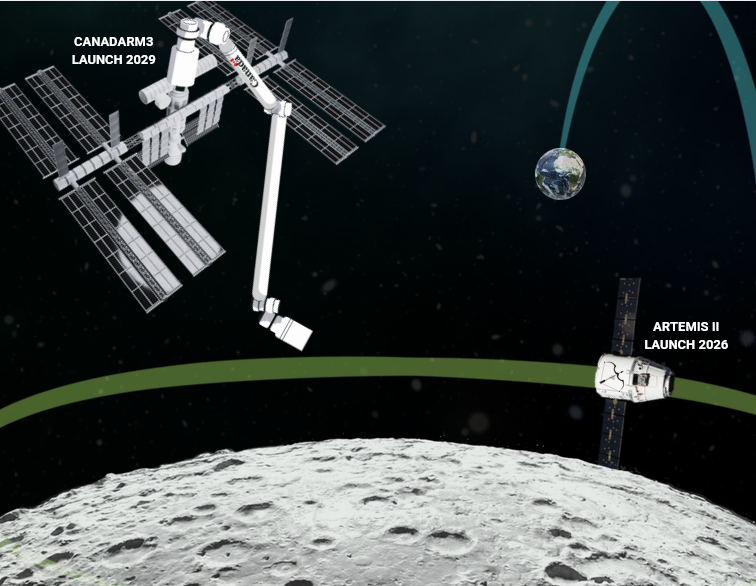}
    \caption{Operational context of Canadarm3 and Artemis II during its lunar fly-by phase, outbound transit in green, and return transit in blue.}    \label{fig:ca-missions}
\end{figure}

As an AI-driven autonomous robotic system, Canadarm3 is designed to operate aboard the Gateway, building on the success of Canadarm2 on the International Space Station (ISS) \cite{oshinowo2012next}. While it will play a pivotal role in future Artemis missions, Artemis II, the first crewed lunar mission since 1972, will mark a significant milestone with astronaut Jeremy Hansen, representing Canada’s first human presence beyond low Earth orbit. This mission will contribute to advancing deep-space exploration and serve as a proving ground for critical technologies \cite{9843277}. Fig.~\ref{fig:ca-missions} illustrates these contributions.

\begin{figure}[ht!]
    \centering
    \includegraphics[width=0.8\linewidth]{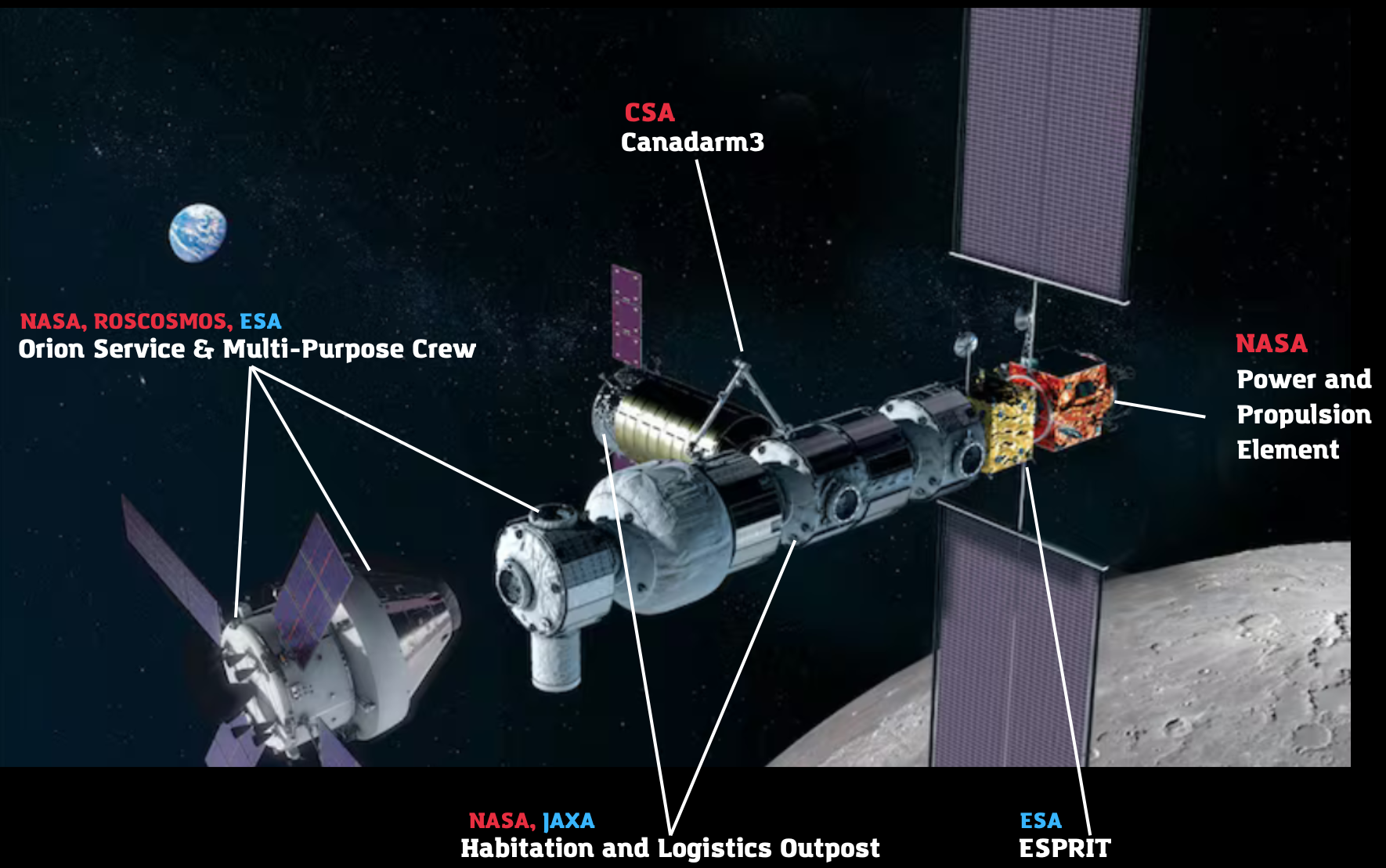}
    \caption{Overview of the Gateway and its systems with each space agency highlighted. The background picture is Orion approaching the Gateway in lunar orbit from \cite{nasa2025lunar}.}
    \label{fig:system-model}
\end{figure}

All Artemis missions are part of a broader objective to establish a sustainable human presence on the Moon. As shown in Fig.~\ref{fig:system-model}, the Gateway is a cornerstone of this next phase of space exploration—an international collaboration that integrates critical technologies. Orbiting the Moon in a near-rectilinear halo orbit (NRHO), it will enable continuous access to the lunar surface and deep-space communication networks, functioning as a highly dynamic system.

\begin{table}[ht!] 
\centering 
\caption{SoS planned on the Gateway.} 
\label{tab:gateway-modules} 
\renewcommand{\arraystretch}{1.1}
\begin{tabular}{|p{1cm}p{3cm}p{3.5cm}|} 
\hline
\textbf{Type} & \textbf{Name} & \textbf{Role} \\ 
\hline 
\hline
\textbf{Core}             & Habitation and Logistics Outpost (HALO)                                      & Serves as the primary crew habitation and logistics support system.             \\ \hline
                                  & International Habitation Module (I-HAB)                                     & Supports various crew activities, providing a habitat environment.              \\ \hline
                                  & ESPRIT Refuelling Module (ESPRIT-ERM) & Provides fuel storage, observation, and cargo management.                       \\ \hline
\textbf{Auxiliary and Power} & Canadarm3       & Assists with cargo handling, docking operations, and general maintenance.       \\ \hline
                                  & HALO Lunar Communication System (ESPRIT-HLCS)                                & Manages telecommunications for the lunar gateway and associated systems.       \\ \hline
                                  & Power and Propulsion Element (PPE)                                          & Supplies power, propulsion, and maneuvering capabilities for the system.       \\ \hline
                                  & Solar arrays                                         & Generates electrical power for station operations.                             \\ \hline
                                  & Fuel storage                                                         & Stores fuel to support spacecraft refueling and station operations.            \\ \hline
                                  & Heat radiators                                               & Dissipates excess heat generated by the onboard systems.                       \\ \hline
\textbf{Docking} & HLS docking port                                                  & Provides interface for docking Human Landing System vehicles.            \\ \hline
                                  & GLS docking port                                                    & Serves as docking interface for Gateway Logistics Services (GLS) spacecraft.   \\ \hline
                                  & Utility docking port                                                      & General-purpose docking interface for various spacecraft or modules.           \\ \hline
                                  & DST docking port                                                   & Docking interface, supporting long-duration missions. \\ \hline
                                  & Return shuttle docking port                                             & Interface for return spacecraft, such as Orion, for crew return to Earth.      \\ \hline\hline 
\end{tabular} 
\end{table}

Table~\ref{tab:gateway-modules} outlines the currently planned \textit{system-of-systems} (SoS) of the Gateway, which itself functions as a highly integrated SoS. Designing such a complex infrastructure requires a holistic approach that balances functionality, economic feasibility, and scalability \cite{doi:10.2514/6.2023-4612}. Additionally, a capability-based design ensures that each SoS supports both its requirements and the broader mission requirements \cite{doi:10.2514/6.2022-1471}.

As a dynamic and evolving SoS, the Gateway must reliably operate within the challenges of the cislunar environment, including long communication delays, deep-space radiation, and phased deployment constraints. Ensuring the resilience of its interconnected SoS is paramount, as failures in any could directly impact astronaut safety and mission success. The Artemis missions, particularly Artemis II, provide an opportunity to validate and refine these systems under real-world conditions, but significant security challenges remain \cite{10115750}. Lessons learned from Canadarm2’s long-term operation on the ISS underscore the importance of anticipating vulnerabilities in interconnected space systems \cite{1303381, Braithwaite2021Canadarm2}. While existing efforts outline broad security measures for cislunar operations \cite{10592290, 10136883}, many remain high-level, necessitating further refinement.

\subsection{Insights from Audit Findings}
\label{related}

A recent audit conducted by the Government Accountability Office (GAO) assessed NASA’s cybersecurity posture, specifically focusing on the Orion Multi-Purpose Crew Vehicle (OMCV), which can dock on the Gateway, and PPE \cite{gao2024cybersecurity}. As the primary crewed spacecraft, OMCV requires stringent security measures to prevent unauthorized access and interference with its positioning and timing systems. It employs a segmentation strategy to isolate its systems. Meanwhile, the PPE focuses on securing command and communication channels. However, the audit highlighted inconsistencies in cybersecurity enforcement, and implementation timelines across these systems. Additionally, even when individual security measures are in place, the lack of overarching integration standards creates gaps in how security is enforced across interconnected systems. 

\subsection{New Standard}

Building on the GAO audit findings, we identify additional gaps in the current evaluation process. Specifically, existing assessments tend to focus on NASA-led SoS and do not fully address the integration of externally developed systems, a critical consideration for the Gateway’s modular and multi-agency architecture. Moreover, according to Gateway specifications \cite{NASA_2019}, the system is classified as a high potential impact, yet security requirements appear to be primarily embedded within the communications domain, one of nine areas. This narrow focus raises concerns, as robust security must extend beyond communications. These concerns are manifested in the following characteristics of the Gateway’s security design:

\begin{itemize}
    \item \textbf{Interdependencies and Cross-System Vulnerabilities}: The interconnected nature of the Gateway's systems means that security risks are not confined to individual systems—vulnerabilities in one can cascade, affecting the integrity of the entire infrastructure.
    \item \textbf{Long-Term Security in a Dynamic Ecosystem}: Given the Gateway’s planned evolution over decades, security measures must account for future system integrations, ensuring adaptability and resilience as new SoS are added.
    \item \textbf{Standardization Across Modules}: With contributions from multiple agencies, maintaining a consistent security architecture is critical. Standardized protocols must be established to ensure interoperability, resilience, and a uniform security baseline across all systems.
    \item \textbf{Sustainability Beyond Deployment}: Security is not a one-time implementation—it must be adaptable to future threats. Systems planned for launch years from now must integrate evolving security controls, necessitating an architecture that supports continuous updates.
\end{itemize}

These complexities underscore the need for a comprehensive security approach, which can be addressed through the \textit{secure-by-component} approach, as part of the forthcoming Space System Cybersecurity standard (also discussed in \cite{10592289,10850060}). The standard ensures that security controls are implemented at the component level, allowing for effective management of security across evolving subsystems and interconnected systems. By embedding security controls within individual components, it ensures that protections remain consistent and resilient when systems are integrated. To achieve this, the standard follows a structured series of steps as follows: \begin{enumerate*}[label=\textbf{(Step \arabic*)}]\item Define engineering scope \item Decompose systems into low-level components, \item Identify attack surfaces \item Identify attack vectors \item Identify \textit{secure-by-design} principles \item Redesign vulnerable components \item Enounciate security requirements \end{enumerate*}. In the following, we will demonstrate how to design secure components by evaluating the two earliest systems, OMCV and PPE. While several frameworks can be used to identify attack vectors (\textbf{Step 4}) and countermeasures (\textbf{Step 5}), we have chosen to utilize the Aerospace Corporation's SPARTA framework \cite{aerospace2023sparta} for the remainder of the paper.

\begin{figure*}[b!]
    \centering
    \resizebox{0.85\linewidth}{!}{
    \begin{tikzpicture}[scale=0.1,font=\large, node distance=2cm, 
    every node/.style={draw, rounded corners, text centered, minimum height=1cm, draw=none, font=\large},
    arrow/.style={thick,->,>=stealth},
    greybox/.style={fill=gray!20, text width=9cm, align=justify, inner sep=5mm, font=\large},
    circlelabel/.style={fill=gray!90, text=black, font=\small\bfseries, inner sep=1mm, minimum size=1mm, shape=circle} 
]
    
    \node [draw=black,minimum width=4.5cm,minimum height=0.85cm] (obc) {Onboard Computer};
    \node [draw=black,minimum width=4.5cm,minimum height=0.85cm, right=0.32cm of obc] (antenna) {Antenna};
    \node [draw=black,minimum width=4.5cm,minimum height=0.85cm, below=0.32cm of obc] (data_processing) {Data Processing/Storage};
    \node [draw=black,minimum width=4.5cm,minimum height=0.85cm, right=0.32cm of data_processing] (signal_processing) {Signal Processing};

    \node[fit=(obc) (signal_processing), draw=blue, minimum height=5cm, minimum width=10cm] (fit) {};
    \node[anchor=south] at (fit.north) (cdh) {C\&DH};

    \node [draw=black, minimum width=7cm, minimum height=0.85cm] (fec) at ([xshift=90cm, yshift=0.3cm] antenna) {Forward Error Correction codes};
    \node [draw=black, minimum width=7cm, minimum height=0.85cm, below=0.32cm of fec] (arq) {Automatic Repeat Request Strategies};
    \node [draw=black, minimum width=7cm, minimum height=0.85cm, below=0.32cm of arq] (eda) {Error detection algorithms};
    \node [draw=white, minimum width=2cm, minimum height=0.85cm, below=0.32cm of eda] (empty) {};

    \node[fit=(fec) (empty), draw=purple, minimum height=5cm, minimum width=7.7cm, right=2.6cm of fit] (new_fit) {};
    \node[anchor=south] at (new_fit.north) (ecc) {ECC};

    \path (fit.east) coordinate (A);
    \path (new_fit.west) coordinate (B);
    \path (fit.east) coordinate (C);
    \path (new_fit.west) coordinate (D);
    
    \draw[->, thick] ([yshift=70pt]A) -- ([yshift=70pt]B) node[pos=0.1, right, yshift=10pt, xshift=-6pt, circlelabel] {1};
    \draw[draw=none] ([yshift=70pt]A) -- ([yshift=70pt]B)  node[pos=0.9, right, yshift=10pt, xshift=-10pt, circlelabel] {2};
    
    \draw[->, thick] ([yshift=-70pt]D) -- ([yshift=-70pt]C)
    node[pos=0.1, right, yshift=-10pt, xshift=-10pt, circlelabel] {3};
    \draw[draw=none] ([yshift=-70pt]D) -- ([yshift=-70pt]C)
    node[pos=0.9, right, yshift=-10pt, xshift=-6pt, circlelabel] {4};
    \end{tikzpicture}
    
    \begin{tikzpicture}[
        font=\large, greybox/.style={draw=none, fill=white, align=left, minimum width=4.5cm, text height=1.5ex, text depth=0.5ex, inner sep=2pt, font=\large, text width=12cm},
        circlelabel/.style={fill=gray!90, text=black, font=\large\bfseries, inner sep=2pt, shape=circle},
        bigbox/.style={draw=none, fill=gray!50, inner sep=4pt},
        node distance=2mm and 2mm
    ]
        \node[circlelabel, yshift=70pt, xshift=40pt] (step1) {1};
        \node[greybox, right=of step1, anchor=west] (desc1) {Data generated by C\&DH needs to be transmitted to other subsystems.};
        
        \node[circlelabel, below=of step1, yshift=-10pt] (step2) {2};
        \node[greybox, right=of step2, anchor=west] (desc2) {Data encoded at ECC uses Forward Error Correction to detect and correct errors during transmission.};
        
        \node[circlelabel, below=of step2, yshift=-10pt] (step3) {3};
        \node[greybox, right=of step3, anchor=west] (desc3) {Data received at the ECC is checked by the Error Detection Algorithm upon receipt. If errors are found, the subsystem attempts to correct them. If correction fails, Automatic Repeat Request is triggered to request a retransmission.};
    
        \node[circlelabel, below=of step3, yshift=-20pt] (step4) {4};
        \node[greybox, right=of step4, anchor=west] (desc4) {Data stored at C\&DH in Data Processing/Storage once corrected can be used.};

    \node at (1.5, 3.2) {\texttt{Flow}};
    \node at (1.5, 6.2) {\texttt{Type}};
    \node at (0.7, 4.7) {};
    \node[draw=purple, minimum width=0.1cm, minimum height=0.1cm, yshift=70pt,rounded corners] (linknode) at (step1.south) {};
    \node[right=of linknode] {Associated with link segment};
    
    \node[draw=blue, minimum width=0.1cm, minimum height=0.1cm, yshift=-10pt,rounded corners](spacenode) at (linknode.south) {};
    \node[right=of spacenode] {Associated with space segment};

     \node[draw=black, minimum width=0.1cm, minimum height=0.1cm, yshift=30pt,rounded corners](componentnode) at (spacenode.south) {};
    \node[right=of componentnode] {Component};
    \end{tikzpicture}
    }
    \caption{Parallel data flows between C\&DH subsystem and ECC.}
\label{fig:decomposition}
\end{figure*}

\begin{table*}[b!]
    \renewcommand{\arraystretch}{1}
    \centering
    \caption{C\&DH subsystem with associated ECC component: security analysis.}
    \resizebox{\textwidth}{!}{
    \begin{tabular}{|p{2cm}|p{3cm}|p{3cm}|p{3.5cm}|p{4cm}|}
        \hline
        \textbf{C\&DH component} & \textbf{Associated ECC component} & \textbf{Attack surface perimeter} & \textbf{Threat techniques} &  \textbf{Security controls} \\ \hline \hline
       
        Onboard Computer & 
        \begin{itemize}[left=0pt, itemsep=0pt]
            \item FEC for commands/telemetry
        \end{itemize} & 
        \begin{itemize}[left=0pt, itemsep=0pt]
            \item \textbf{Input:} Command interfaces
            \item \textbf{Output:} Processed telemetry
            \item \textbf{Dependency:} Firmware \& software libraries
        \end{itemize} & 
        \begin{itemize}[left=0pt, itemsep=0pt, label={}]
            \item EX-0005 Exploit Hardware/Firmware Corruption
            \item EX-0009 Exploit Code Flaws
            \item EX-0012 Modify On-Board Values
        \end{itemize}
        & 
        \begin{itemize}[left=0pt, itemsep=0pt, label={}]
            \item \textbf{$\bigstar$$\clubsuit$}CM0038 Segmentation 
            \item \textbf{$\bigstar$}CM0014 Secure boot 
            \item CM0043 Backdoor Commands 
            \item \textbf{$\bigstar$}CM0045 Error Detection and Correcting Memory 
        \end{itemize}
        \\ \hline
       
        Data Processing/Storage & 
        \begin{itemize}[left=0pt, itemsep=0pt]
            \item Error detection algorithms
            \item ARQ for corrupted data retrieval
        \end{itemize} & 
        \begin{itemize}[left=0pt, itemsep=0pt]
            \item \textbf{Input:} Data streams
            \item \textbf{Output:} Stored/retrieved data
            \item \textbf{Dependency:} Storage media, I/O processes
        \end{itemize} & 
        \begin{itemize}[left=0pt, itemsep=0pt, label={}]
            \item EX-0012 Modify On-Board Values
            \item EX-0013 Flooding 
            \item EXF-0001 Replay 
        \end{itemize}
        & 
        \begin{itemize}[left=0pt, itemsep=0pt, label={}]
            \item CM0036 Session Termination 
            \item \textbf{$\bigstar$$\clubsuit$}CM0034 Monitor Critical Telemetry Points 
            \item CM0033 Relay Protection 
            \item \textbf{$\bigstar$$\clubsuit$}CM0039 Least Privilege 
            \item CM0056 Data Backup
        \end{itemize}
        \\ \hline       
       
        Antennas & 
        \begin{itemize}[left=0pt, itemsep=0pt]
            \item FEC in signal modulation/demodulation
        \end{itemize} & 
        \begin{itemize}[left=0pt, itemsep=0pt]
            \item \textbf{Input:} Incoming RF signals
            \item \textbf{Output:} Transmitted signals
            \item \textbf{Dependency:} Signal modulation systems
        \end{itemize} & 
        \begin{itemize}[left=0pt, itemsep=0pt, label={}]
            \item EX-0016 Jamming 
            \item EXF-0003 Eavesdropping 
            \item EXF-0006 Modify Communications Configuration 
        \end{itemize}
        & 
        \begin{itemize}[left=0pt, itemsep=0pt, label={}]
            \item \textbf{$\clubsuit$}CM0029 TRANSEC 
            \item CM0083 Antenna Nulling and Adaptive Filtering 
            \item CM0085 Electromagnetic Shielding 
        \end{itemize}
        \\ \hline
       
        Signal Processing Unit & 
        \begin{itemize}[left=0pt, itemsep=0pt]
            \item Error detection and correction in signal decoding
        \end{itemize} & 
        \begin{itemize}[left=0pt, itemsep=0pt]
            \item \textbf{Input:} Amplified RF signals
            \item \textbf{Output:} Decoded data
            \item \textbf{Dependency:} Amplifiers, decoders
        \end{itemize} & 
        \begin{itemize}[left=0pt, itemsep=0pt, label={}]
            \item EX-0015 Side-Channel Attack
            \item EEX-0014 Spoofing 
            \item EXF-0006 Modify Communications Configuration
        \end{itemize}
        & 
        \begin{itemize}[left=0pt, itemsep=0pt, label={}]
            \item CM0064 Dual Layer Protection  
            \item $\clubsuit$CM0028 Tamper Protection 
            \item CM0061 Power Masking 
        \end{itemize}
        \\ \hline
    \end{tabular}}
    \label{tab:cdh}
\end{table*}

\section{Securing Early Systems}
\label{theo}
In the first phase of Gateway deployment, PPE and HALO will be integrated and launched together by 2027, with the OMCV launching a year later for the crewed assembly of the I-HAB module. Thus, we focus on analyzing the PPE and OMCV which establish the initial infrastructure, focusing on:
\begin{enumerate}[label=\textbf{(Focus \arabic*)}, left=0pt, labelsep=1em]
    \item A single engineering scope that avoids one hazard that can arise from both intentional and unintentional sources.
    \item Three key segments—space, integration, and link—which are integral to data exchange, command transmission, and operational coordination between mission elements \textit{in} space. This aligns with the objectives of the audit, such as maintaining command authority and detecting interference. 
    \item One subsystem per segment which is shared in both SoS to allow for a comparative analysis while contextualizing the findings for each. 
    \item The two shared subsystems perform critical functions of data distribution and error management, supporting reliable and resilient communication, even in the event of a hazard.
    \item Set of the most critical security controls that apply to each SoS. 
\end{enumerate}

\subsection{Common Blocks for Analysis}
\noindent We begin by defining one critical hazard associated with these systems: the loss of control of the space vehicle (\textbf{Focus 1} - \textbf{Step 1}) which can lead to mission failure, loss of spacecraft, or, in the worst cases, the loss of crew members. We select two subsystems present in both systems: the communication and data handling (C\&DH) subsystem in the space segment, which ensures the accurate and secure distribution of mission-critical data, and the error correction code (ECC) in the link segment, which plays a key role in managing errors and ensuring data integrity during transmission (\textbf{Focus 2\(\rightarrow \)4}). Fig. \ref{fig:decomposition} illustrates the data flows between these two subsystems. With this, we provide their breakdown into their components, attack surfaces, attack techniques, and associated countermeasures, as shown in Table \ref{tab:cdh} (\textbf{Steps 2\(\rightarrow \)5}).  The most critical countermeasures for OMCV are marked with (\textbf{$\bigstar$}), and for PPE with (\textbf{$\clubsuit$}) in the table. They will be used in the redesign process and in defining corresponding security requirements (\textbf{Steps 6\(\rightarrow \)7}), which will be discussed for each system (\textbf{Focus 4\(\rightarrow \)5}). They ensure secure connection and communication in \begin{enumerate*}[label=(\roman*)] \item \textit{internal}, within each system and \item \textit{external}, with other systems in the Gateway. \end{enumerate*}

\subsection{Orion Multi-Purpose Crew Vehicle}
\noindent Given its crewed nature, a primary countermeasure is Segmentation (CM0038) to isolate critical subsystems and limit impact. Additionally, Error Detection and Correcting Memory (CM0045) mitigates data corruption from cosmic radiation and system faults. A corresponding security requirement is:
\begin{quote}
\textit{  
\small
    OBC \textsc{SHALL} implement Error Detection and Correcting Memory (CM0045) to ensure the integrity of telemetry and command data, safeguarding against potential data corruption caused by space environment interference.
}
\end{quote}

\subsection{Gateway Power and Propulsion Element}
\noindent While similar principles apply, PPE requires TRANSEC (CM0029), which ensures that the communication is encrypted, making it difficult for unauthorized parties to intercept or manipulate the data. Additionally, Monitoring Critical Telemetry Points (CM0034) provides real-time surveillance of key data points to detect anomalies. A security requirement for PPE is: 
\begin{quote}
\textit{    
\small
    Data processing/Storage \textsc{SHALL} implement TRANSEC (CM0029) to ensure the confidentiality and integrity of communication signals.}
\end{quote}

\noindent Notably, the controls identified align with \cite{gao2024cybersecurity}. While the audit may not disclose all controls, our methodology demonstrates how these controls can be identified and examined.

\subsection{Integration Considerations}
\noindent The analysis of individual SoS reveals specific security requirements, but it is crucial to consider their integration within the broader Gateway architecture. To ensure a resilient and standardized security posture, a comprehensive baseline of security requirements must be established, incorporating all critical security controls identified. In addition to these baseline requirements, the unique operation of the C\&DH module—functioning both independently and as a shared resource \cite{NASA_2019}—warrants the inclusion of additional controls, such as Robust Fault Management (CM0042) and Shared Resource Leakage Protection (CM0040). These controls are essential for safeguarding the system’s integrity and fault resilience across both individual SoS and the integrated SoS.

\noindent While the baseline security requirements ensure consistency and standardization, they do not preclude the application of additional, SoS-specific controls. Specifically, modules such as the docking and core systems, which involve crew presence, require the highest level of resilience. These SoS would also be fortified with stringent measures, such as Physical Security Controls (CM0053) and the Two-Person Rule (CM0054), to further enhance the protection of crew and critical operations. These additional security measures complement the standardized baseline controls while addressing the unique needs of crewed modules.

\section{Canadarm3 Integration}
\label{canadarm3}

\begin{figure}[t!]
\centering
\resizebox{\linewidth}{!}{
\includegraphics[width=\linewidth]{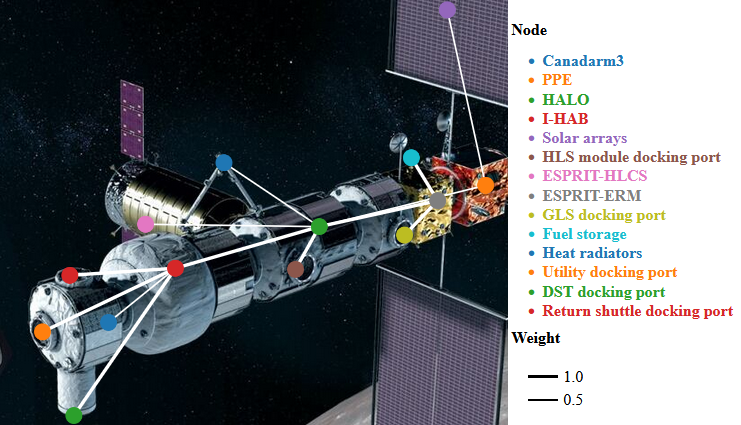}
}
\caption{Graph presentation of Gateway and structural links.}
\label{fig:graph theory-label}
\end{figure}

If we now shift to an operational context where Canadarm3 is integrated alongside other Gateway modules, we propose assessing the effectiveness of security controls in ensuring system resilience. While these controls are designed during the onset, their real-world performance remains critical. One such operational risk is collision with objects, which can lead to loss of control (\textbf{Focus 1} - \textbf{Step 1}). Past incidents highlight the severity of this risk, including the 2021 collision that nearly severed Canadarm2 and the increasing number of avoidance maneuvers—now totaling 33 for the ISS \cite{NASA_ODQN_2023}. Current countermeasures primarily rely on shielding and collision avoidance protocols. However, within the Gateway context, the challenge is exacerbated by reduced situational awareness due to the absence of dedicated Space Domain Awareness (SDA) systems, making collision risks more hazardous.

\textbf{Case Study - Collision Impact Propagation: }
Our analysis takes a cyber-physical approach, as collisions involve both digital manipulation (cyber) and physical consequences. A plausible cause for such an event could stem from the autonomy of Canadarm3, which relies on AI to detect and avoid collisions. Any compromise—such as a software malfunction, cyber attack, or sensor failure—could lead to either an undetected collision or an over-detected one, triggering unnecessary evasive maneuvers or fail-safes.

We use a dataset from MDA, including 330 collision trials, to assess the arm’s collision avoidance capabilities \cite{MDA2021}. While the dataset trains the arm’s AI, its reliance on autonomy introduces vulnerabilities. In the absence of AI-based avoidance, the initial collision probability is 33\%. In the following, we analyze failure propagation from Canadarm3, through the physical connections. We model the Gateway as an undirected weighted graph, as illustrated in Fig.\ref{fig:graph theory-label}. The graph is formally defined as,
\begin{equation}
    G = (V, E, W),
\end{equation}

where $V$ represents the set of nodes, $E$ denotes the set of edges, and $W$ is the weight function \( W: E \to \mathbb{R} \) which assigns values to each edge based on the nature of the connection: direct, fixed physical links (e.g., structural interfaces) are assigned a weight of 1, whereas auxiliary dependencies (e.g., robotic interactions or indirect support mechanisms) are assigned a weight of 0.5.

A failure can propagate from one node to adjacent nodes over time through structural links. The impact on node $v$ from an adjacent node $u$ at time $t$ is given by,\begin{equation}
    I(v,t+1) =  W_{u,v} \cdot I(u,t),
\end{equation}
where $I(v,t)$ represents failure impact. $W_{uv}$ is the weight of the structural link between $u$ and $v$.

In the above, failure spreads without attenuation—once a failure reaches a node, it remains affected. To introduce the attenuation from security controls, we refine the model as follows,
\begin{equation}
    I(v,t+1) = max \Bigg( I(v,t), \alpha  \cdot  \sum_{u \in N(v)} W_{u,v} \cdot I(u,t) \Bigg),
\end{equation}
where $\alpha$ represents a diffusion factor; when $\alpha$ is close to 0, the countermeasures effectively isolate the failure to the node, preventing its diffusion to other parts of the system. 

\begin{figure}[t!]
\centering
\includegraphics[width=0.85\linewidth]{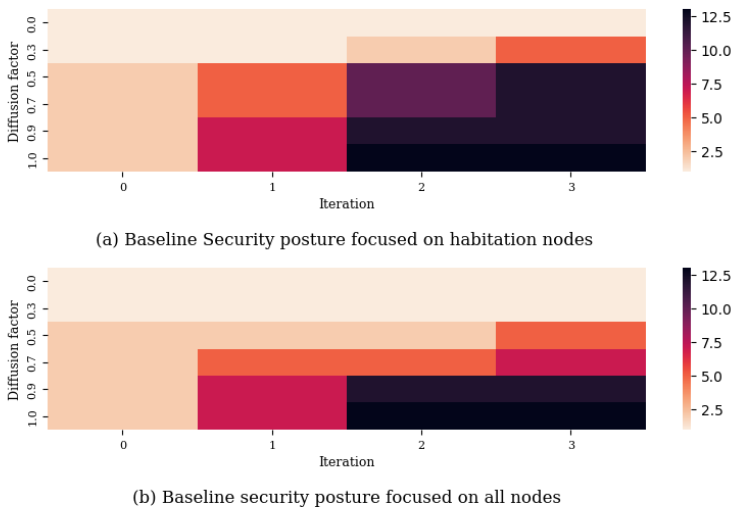}
\caption{Propagation of collision fault.}
\label{fig:prop}
\end{figure}

Fig. \ref{fig:prop} illustrates the results, highlighting the critical role of structural connectivity in failure propagation and the need for consistent countermeasure application. The baseline security strategy, where all critical controls are uniformly applied (b), proves to be the most effective in containing failure spread, emphasizing its importance as the strongest security foundation. In comparison, applying countermeasures solely to habitation modules results in a 70\% increase in the number of affected nodes, even with a low diffusion factor of $<0.5$. Our findings reveal that failures originating from Canadarm3 can propagate to HALO, even through reduced-weight links, and despite increased countermeasures applied to habitation nodes. However, the spread can be significantly reduced when a uniform baseline is applied across all nodes. If failure is not contained, it severely weakens the overall system integrity. This highlights the importance of future architectural designs, where node weighting considers not only functional roles but also structural resilience—identifying which nodes can be isolated or removed during contingency scenarios. To support the insight, we proved that HALO emerges as the most vulnerable node due to its high connectivity and bridges to the rest of the network. While countermeasures help mitigate propagation, complete isolation remains unfeasible, as HALO must remain connected to PPE for crew safety and station stability. Additionally, the Gateway's inability to fully section off compromised areas could pose a significant limitation. Since only PPE and docked spacecraft provide propulsion, containment strategies become critical in failure scenarios.

\section{Conclusion}
\label{conc}

This paper presents an SoS analysis of the Gateway, developing baseline security controls for C\&DH across link, integration, and space segments. It also explores security challenges in Canadarm3 integration, highlighting cross-system vulnerabilities. Future work will assess Canadarm3’s resilience to failure scenarios and extend propagation modeling to include timing factors, evaluating how response and detection delays impact the effectiveness of mitigation measures.
\bibliographystyle{IEEEtran} 
\bibliography{references}

\end{document}